\documentclass[11pt,twoside]{article}
\usepackage{cspm2015}

\resetcounters

\def\figspath{.} 

\begin{document}

\title{Flows in and around active region NOAA12118 observed with the GREGOR
solar telescope and SDO/HMI}

\author{
    M.\ Verma,$^1$
    C.\ Denker,$^1$   
    H.\ Balthasar,$^1$
    C.\ Kuckein,$^1$
    S.J.\ Gonz{\'a}lez Manrique,$^{1,2}$
    M.\ Sobotka,$^{6}$
    N.\ Bello Gonz{\'a}lez,$^{3}$  
    S.\ Hoch,$^{3}$,   
    A.\ Diercke,$^{1,2}$
    P.\ Kummerow,$^{1,2}$   
    T.\ Berkefeld,$^{3}$ 
    M.\ Collados,$^{5}$
    A.\ Feller,$^{4}$
    A.\ Hofmann,$^{1}$
    F.\ Kneer,$^{7}$
    A.\ Lagg,$^{4}$
    J.\ L{\"o}hner-B{\"o}ttcher,$^{3}$
    H.\ Nicklas,$^{7}$ 
    A.\ Pastor Yabar,$^{5, 10}$
    R.\ Schlichenmaier,$^{3}$
    D.\ Schmidt,$^{3, 8}$
    W.\ Schmidt,$^{3}$
    M.\ Schubert,$^{3}$
    M.\ Sigwarth,$^{3}$
    S.K.\ Solanki,$^{4, 9}$
    D.\ Soltau,$^{3}$        
    J.\ Staude,$^{1}$
    K.G.\ Strassmeier,$^{1}$
    R.\ Volkmer,$^{3}$
    O.\ von der L{\"u}he,$^{3}$ and
    T.\ Waldmann$^{3}$    
\affil{%
    $^1$Leibniz-Institut f{\"u}r Astrophysik Potsdam (AIP),
        Germany; \email{mverma@aip.de}}
\affil{%
    $^2$Universit{\"a}t Potsdam,
        Institut f{\"u}r Physik und Astronomie, Germany}
\affil{%
    $^3$Kiepenheuer-Institut f{\"u}r Sonnenphysik, Germany}
\affil{%
    $^4$Max-Planck-Institut f{\"u}r Sonnensystemforschung, Germany}
\affil{%
    $^5$Instituto de Astrof\'{\i}sica de Canarias, Spain}
\affil{%
    $^6$Astronomical Institute, Academy of Sciences of the Czech Republic}
\affil{%
    $^7$Institut f\"ur Astrophysik, Georg-August-Universit\"at G\"ottingen,
        Germany}
\affil{%
    $^8$National Solar Observatory, USA}
\affil{%
    $^9$School of Space Research, Kyung Hee University, Korea}
\affil{%
    $^{10}$Dept. Astrof\'{\i}sica, Universidad de La Laguna, Tenerife, Spain}}

\paperauthor{M.\ Verma}{mverma@aip.de}{}{Leibniz-Institut f{\"u}r 
Astrophysik Potsdam (AIP)}{}{}{}{}{}

\begin{abstract}
Accurate measurements of magnetic and velocity fields in and around solar active 
regions are key to unlocking the mysteries of the formation and the decay of 
sunspots. High spatial resolution image and spectral sequences with a high 
cadence obtained with the GREGOR solar telescope give us an opportunity to 
scrutinize 3-D flow fields with local correlation tracking and imaging 
spectroscopy. We present GREGOR early science data acquired in 2014 
July\,--\,August with the GREGOR Fabry-P\'erot Interferometer and the 
Blue Imaging Channel. Time-series of blue continuum ($\lambda$~450.6 nm) 
images of the small active region NOAA~12118 were restored with the speckle 
masking technique to derive horizontal proper motions and to track the evolution 
of morphological changes. In addition, high-resolution observations are 
discussed in the context of synoptic data from the Solar Dynamics Observatory.
\end{abstract}

\section{Introduction}

With new instruments and improved observing facilities we can enhance our 
understanding of plasma motions surrounding sunspots and their interaction with 
magnetic fields. Current developments in the magneto-hydrodynamics (MHD) 
simulations \citep[e.g.,][]{2014ApJ...785...90R} provide the complete evolution 
of an active region from flux emergence to decay, allowing direct comparison 
with these high-resolution observations. In the growing phase, penumbra form 
rapidly around the sunspot \citep{1998ApJ...507..454L, 2003ApJ...597.1190Y}. The 
appearance of penumbrae is closely linked to inclined magnetic field lines and 
the onset of the Evershed flow. During the growth of the penumbra the umbral 
size seems to remain stable according to \citet{2010A&A...512L...1S}. It is 
common that the growth of sunspots is faster than the decay 
\citep[e.g.,][]{2012A&A...538A.109V}. The stable leading sunspots and irregular 
trailing spots decay with different rates \citep{2002AN....323..342M}. Many 
decay laws have been proposed ranging from a linear decay rate of 
\citet{1963BAICz..14...91B} to a parabolic decay law described by 
\citet{1997SoPh..176..249P}. 

\articlefigure[width=.85\textwidth]{\figspath/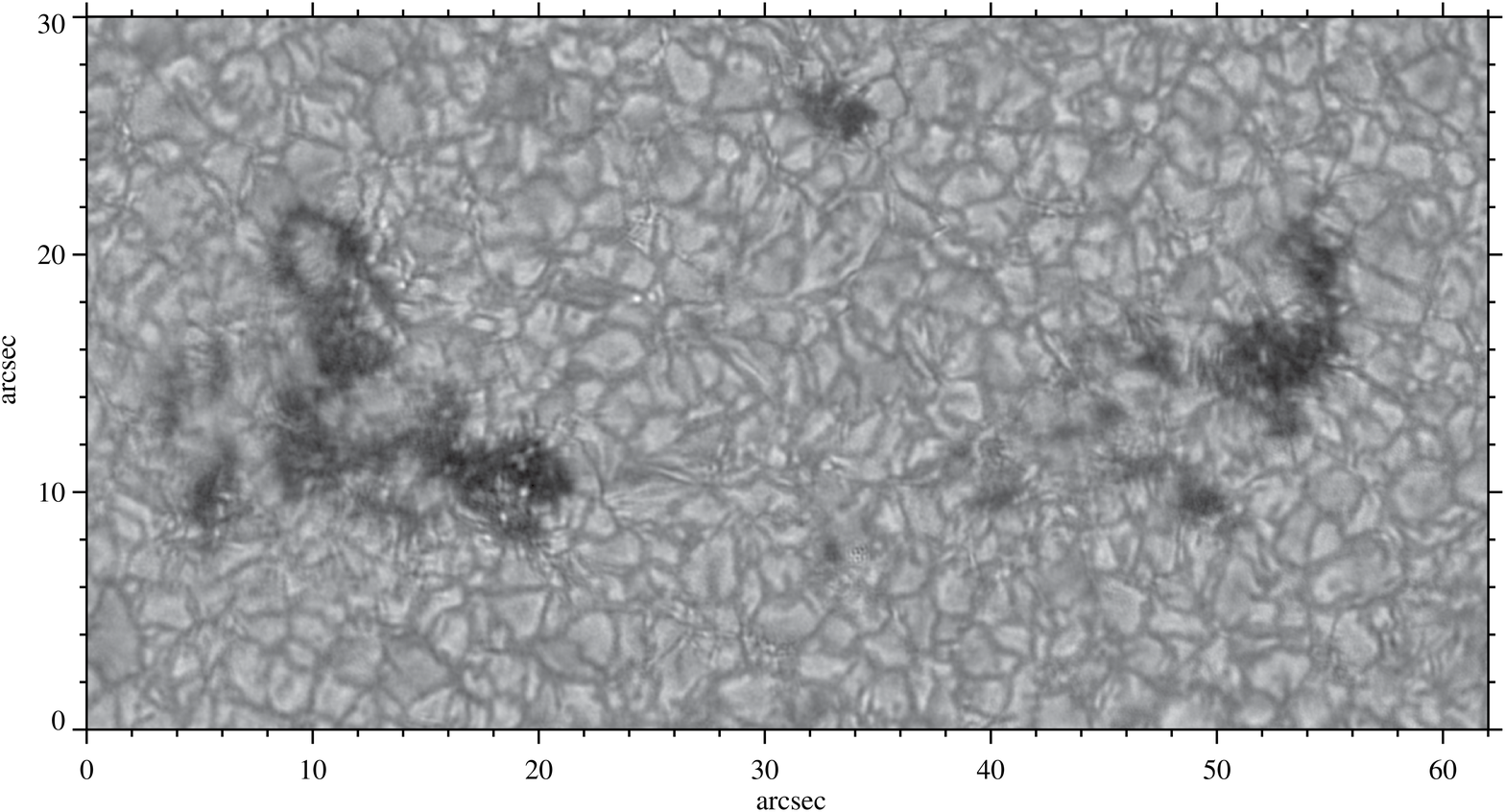}{fig01}{ 
Bipolar active region NOAA~12118 observed in a blue continuum image of 
GREGOR/BIC on 2014 July~18. The image was rotated to match the orientation 
of the region in SDO images.}

The flow structure around sunspots and active region plays significant role in 
both growth and decay processes. Magnetic flux from the sunspot is removed via 
the interaction of penumbral filaments and the surrounding granulation 
\citep{2008ApJ...686.1447K}. Unipolar moving magnetic features 
\citep[e.g.,][]{2005ApJ...635..659H}, which are usually observed around decaying 
sunspots \citep{1973SoPh...28...61H} and pores \citep{2012A&A...538A.109V}, are 
supposed to transport flux to surrounding supergranular cell boundaries. Hence, 
detailed knowledge of flow fields in and around solar magnetic features at 
various evolutionary stages is expected to provide insight into the growth and 
decay processes of sunspots. In this work based on GREGOR data, we present the 
horizontal proper motions around an active region, which is in its growth phase 
with continuous flux emergence.

\section{Observations and data reduction}

In July\,--\,August 2014 we had carried out a 50-day-long ``early science'' 
campaign at the 1.5-meter GREGOR solar telescope \citep{2012AN....333..810D, 
2012AN....333..796S}, which resulted in 30~days of observations with the Blue 
Imaging Channel (BIC) and the GREGOR Fabry-P\'erot Interferometer 
\citep[GFPI,][]{2010SPIE.7735E..6MD, 2012AN....333..880P}. Here, we present 
time-series of blue continuum ($\lambda$~450.6 nm) images of the bipolar active 
region NOAA~12118 observed on 2014 July~18 starting at 07:56~UT. The four 
time-series of 30~min each were observed with 60 image sequences containing 80 
images each with an exposure time of 4~ms. The images have a size of 
2160$\times$2672~pixels with an image scale of 0.035~\arcsec~pixel$^{-1}$. The 
resulting field-of-view (FOV) of $75^{\prime\prime} \times 93^{\prime\prime}$ 
fully contains the active region. The blue continuum images were reconstructed 
using the Kiepenheuer Institute Speckle Interferometry Package 
\citep[KISIP,][]{2008A&A...488..375W, 2008SPIE.7019E..1EW}. Figure~\ref{fig01} 
shows an image from the blue continuum time-series. A region-of-interest (ROI) 
was extracted from the image and rotated to remove the image rotation introduced 
by the altitude-azimuth mount of the GREGOR telescope.

\articlefigure[width=.85\textwidth]{\figspath/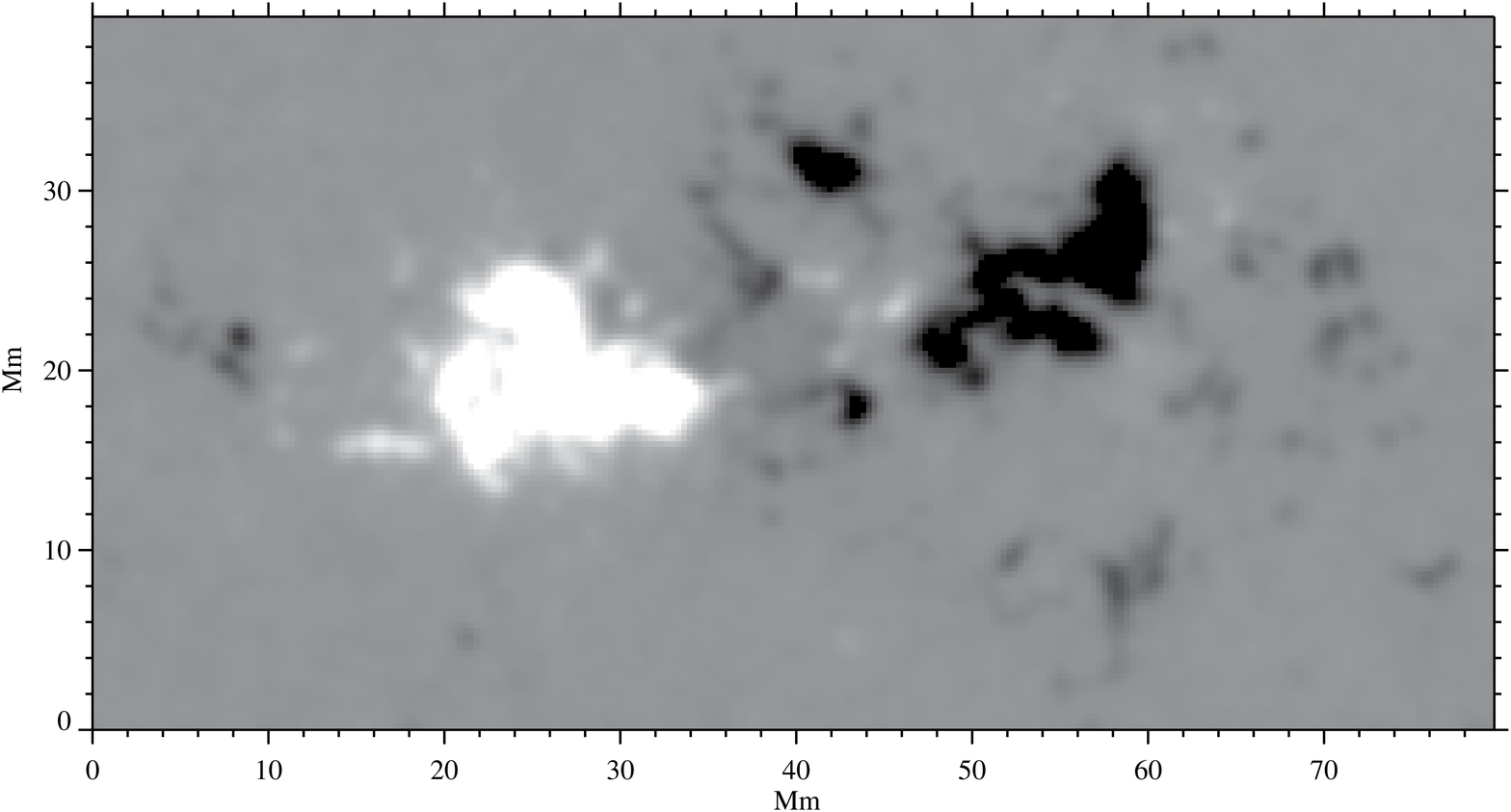}{fig02}{ 
Active region NOAA~12118 as observed in a SDO/HMI LOS magnetogram. The 
magnetogram is displayed between $\pm 500$~G.}

We complemented GREGOR observations with the synoptic observations from the 
Solar Dynamics Observatory \citep[SDO,][]{2012SoPh..275..207S}. SDO's 
Helioseismic and Magnetic Imager \citep[HMI,][]{2012SoPh..275..229S, 
2012SoPh..275..285C, 2012SoPh..275..261W} provided line-of-sight (LOS) 
magnetograms and continuum images covering the entire evolution of the active 
region, i.e., from its emergence to decay. The time-cadence of 45~s allowed us 
to use a time-series of LOS magnetograms for computing horizontal proper 
motions. Apart from basic calibration steps, magnetograms were corrected for 
geometrical foreshortening, after which a pixel represented 360~km on the solar 
surface. Figure~\ref{fig02} shows one of the magnetograms. The 40~Mm $\times$ 
80~Mm ROI was extracted from the full-disk magnetogram. 

Active region NOAA~12118 appeared on the solar disk on 2014 July~17 around 
15:00~UT. By the time of the GREGOR observations the region was approaching its 
maximum growth. However, the central part of the region was still a site for 
continuous flux emergence. The region was bipolar and categorized as an $\alpha 
\beta$-group. The negative-polarity leading and positive-polarity trailing parts 
were both conglomerations of pores, which never developed penumbrae. The region 
started to decay on 2014 July 19. Atypically, the leading pore decayed first 
\citep{1963BAICz..14...91B}, and the region had significantly decayed by 2014 
July 22.

\articlefiguretwo{\figspath/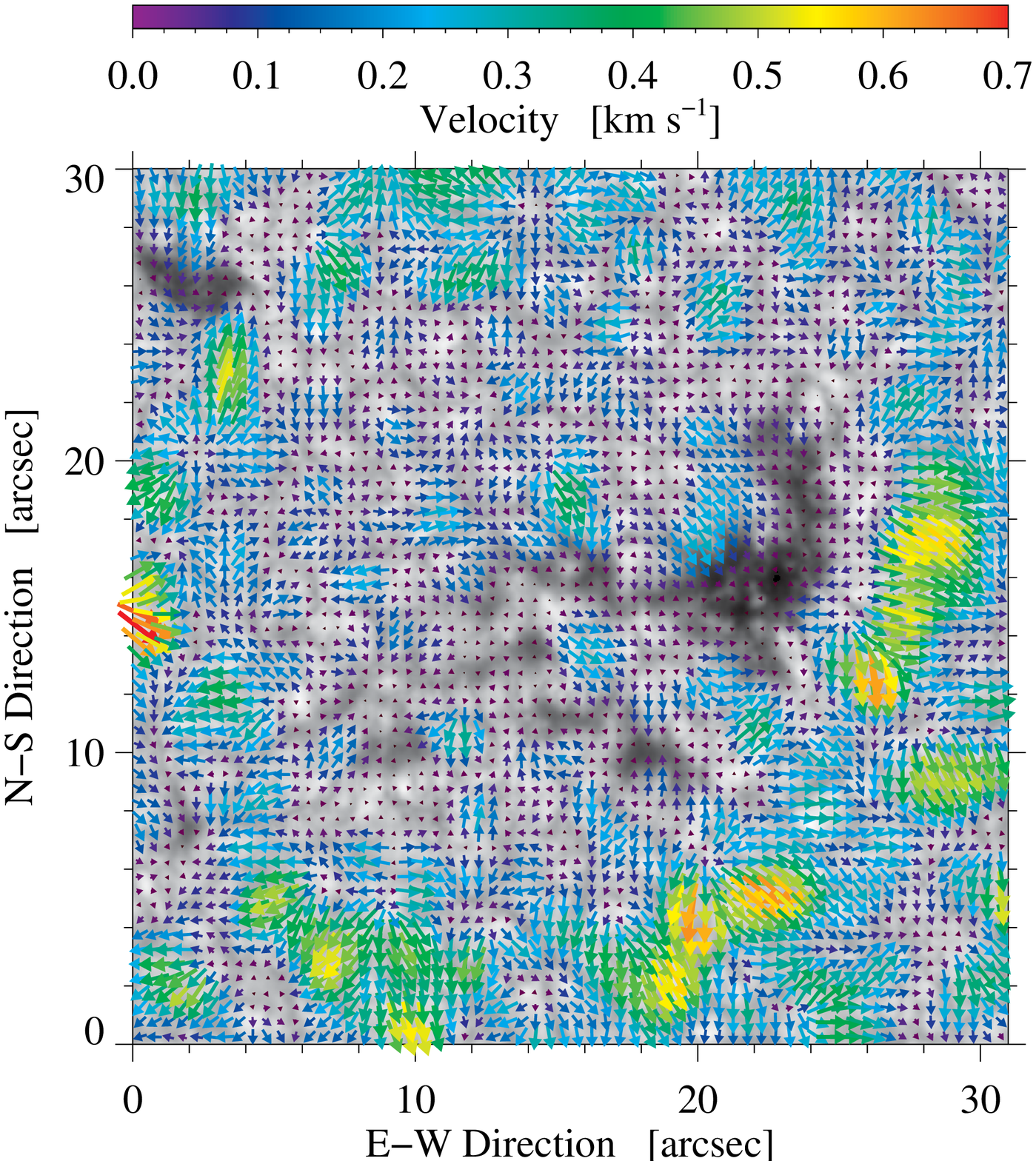} 
{\figspath/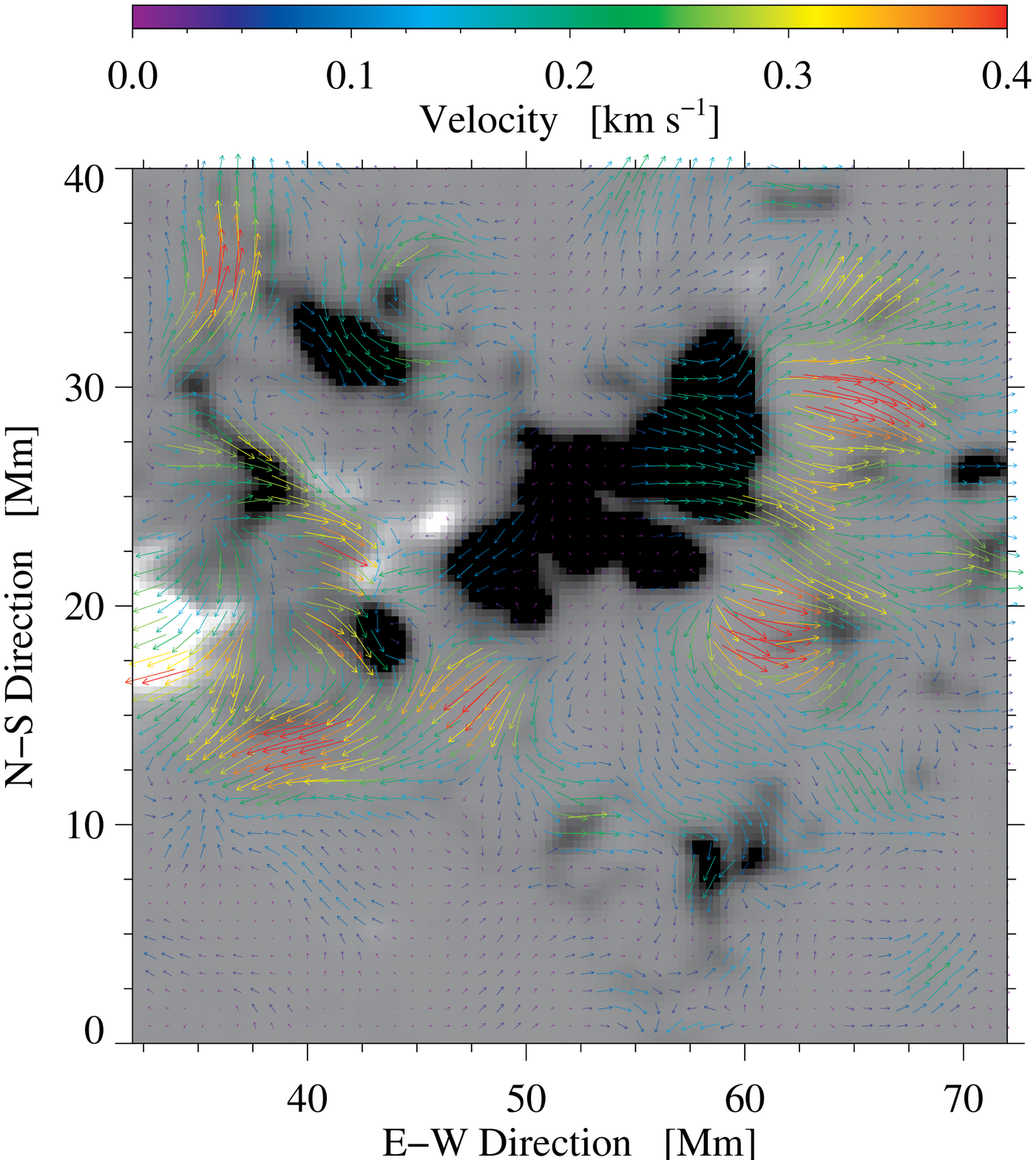}{fig03}{Detailed view of horizontal 
proper motions around the leading part of the active region computed using LCT 
(\textit{left}) and DAVE (\textit{right}). The coordinates refer to 
Figs.~\ref{fig01} and \ref{fig02}, respectively. In these images one~arcsec 
corresponds to 0.78~Mm. The color figure can be found in the electronic 
version.}

Horizontal proper motions using GREGOR continuum images were computed with the 
Local Correlation Tracking (LCT) code of \citet{2011A&A...529A.153V}. The code 
was adapted for high-resolution data. Images were aligned and the signature of 
five-minute oscillations was removed. Cross-correlations were computed over $48 
\times 48$-pixel image tiles with a Gaussian kernel having a FWHM = 1200~km, 
corresponding to the typical size of a granule. The time cadence was $\Delta t = 
60$~s, and the flow maps were averaged over $\Delta T = 30$~min. Corresponding 
to GREGOR data we chose a 2-hour time-series of LOS magnetograms starting at 
08:00 UT applying the Differential Affine Velocity Estimator 
\citep[DAVE,][]{2005ApJ...632L..53S, 2006ApJ...646.1358S} to calculate the 
horizontal plasma velocities. We computed horizontal proper motions for all four 
time-series. However, considering the scope of this conference contribution we 
present preliminary results only for the second time-series, i.e., the 30-minute 
time-series starting 08:36~UT. 

\section{Results}


To infer the proper motions in continuum images and magnetograms we used LCT and 
DAVE, respectively. These methods were applied to the full FOV. However, here we 
focus only on the leading pore of the active region (Fig.~\ref{fig03}).

The observed active region was bipolar, with leading negative polarity. In the 
DAVE map (right-panel Fig.~\ref{fig03}) there are strong outward motions along 
the border of the dominant pore. Inside the pores the horizontal velocities are 
lower and uniform. However, at coordinates (45~Mm, 22.5~Mm) in a region with 
mixed polarity the proper motion is multidirectional. At coordinates (47.5~Mm, 
35~Mm) weak swirling motion is visible. The maximum DAVE velocity in the ROI 
reached up to 0.4~km~s$^{-1}$. 

The LCT map provides horizontal proper motions by tracking on photospheric 
features in continuum images. In the 30-minute averaged map of 30\arcsec\ 
$\times$ 30\arcsec, we see high velocities in the surroundings of the pore 
giving an impression of moat flow. However, as already mentioned, none of the 
pores in the active region developed a penumbra, i.e., no Evershed flow can 
develop. In addition, many small scale diverging centers are visible, e.g., at 
coordinates (2\arcsec, 20\arcsec) and (14\arcsec, 29\arcsec). The LCT method 
tracks photospheric contrast features (granules, penumbral filaments, umbral 
dots, etc.) and does not per se rely on the presence of magnetic fields as in 
the case of DAVE. The maximum velocities achieved in the LCT map reach up to 
0.7~km~s$^{-1}$. Already visual inspection clearly reveals differences in 
velocities computed by DAVE and LCT. This discrepancy points to differences in 
the underlying physics used in both methods, but also the disparate spatial 
resolution has to be considered.

\articlefigure[width=.89\textwidth]{\figspath/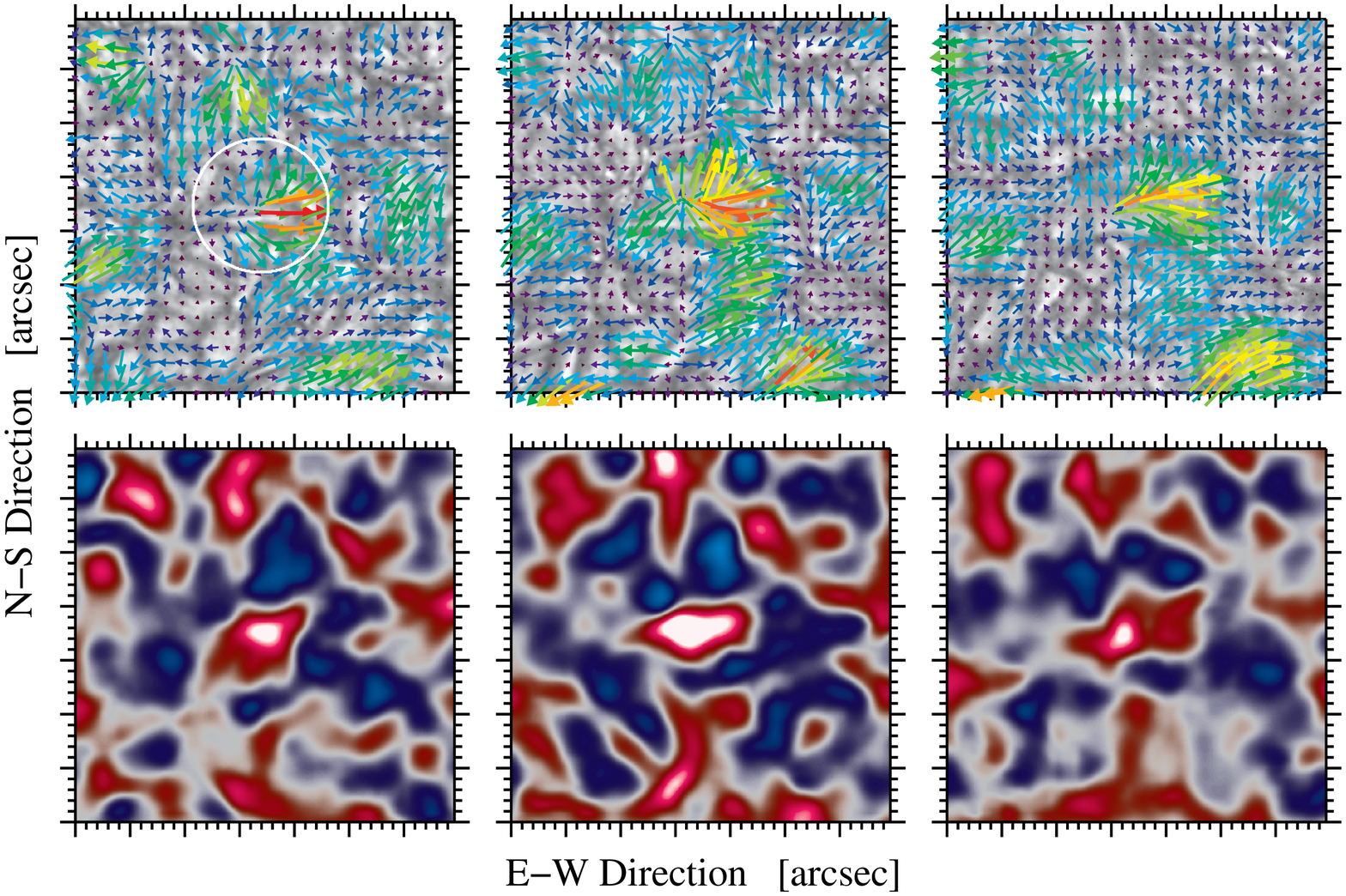}{fig04}
{LCT velocities (\textit{top}) and divergence (\textit{bottom}) for one of the 
rapidly expanding granules marked by white circle over the period of 30 minutes. 
The FOV is 14\arcsec $\times$ 14\arcsec. The rainbow-colored averaged LCT 
vectors are superposed on the 10-minute average continuum images of GREGOR, 
which capture the evolution of the rapidly expanding granule. Red colors 
represents divergence and blue corresponds to convergence. Divergence maps are 
scaled between $\pm 5 \times 10^{-2}$~s$^{-1}$. The color figure can be found in 
the electronic version.} 

The high spatial and temporal resolution of the GREGOR data allows us to follow 
horizontal proper motions of small-scale features such as \textit{rapidly 
expanding granules}. In this active region, we see many granules which are 
larger and more elongated than normal granules. These rapidly expanding granules 
could be the signature of continuous flux emergence. One example of these 
expanding granules is shown in Fig.~\ref{fig04}. We show an ROI of 14\arcsec 
$\times$ 14\arcsec centered on the granule. The horizontal flow maps are 
averaged over 10~min, and the background is the corresponding averaged blue 
continuum image. The center of the granule has a positive divergence from the 
start (as seen in the first panel of Fig.~\ref{fig04}), which becomes highest in 
the next 10~minutes, while in the last 10~minutes it decreases. The LCT velocity 
vectors traced the full areal extent of the rapidly expanding granules.

\section{Conclusion}

The GREGOR high-resolution images and SDO LOS magnetograms furnished information 
about the evolution of horizontal flow fields around NOAA~12118. In the LCT and 
DAVE flow maps outward flows along the border of the leading pore were present. 
In addition, LCT captured the diverging proper motions of rapidly expanding 
granules. The presence of these granules in the region can be related to 
continuous flux emergence. In the future, the combination of spectropolarimetric 
observations of the GFPI and the GREGOR Infrared Spectrograph 
\citep[GRIS,][]{2012AN....333..872C} along with high-resolution imaging will 
provide the opportunity to follow changes in flows as well as magnetic fields. 
The multi-wavelength and multi-instrument setup is necessary to understand the 
complex process of spot formation and decay.

\acknowledgements The 1.5-meter GREGOR solar telescope was build by a German 
consortium under the leadership of the Kiepenheuer-Institut f\"ur Sonnenphysik 
in Freiburg with the Leibniz-Institut f\"ur Astrophysik Potsdam, the Institut 
f\"ur Astrophysik G\"ottingen, and the Max-Planck-Institut f\"ur 
Sonnensystemforschung in G\"ottingen as partners, and with contributions by the 
Instituto de Astrof\'{\i}sica de Canarias and the Astronomical Institute of the 
Academy of Sciences of the Czech Republic. SDO HMI and AIA data are provided by 
the Joint Science Operations Center -- Science Data Processing. MS is supported 
by the Czech Science Foundation under the grant 14-0338S. CD have been supported 
by grant DE 787/3-1 of the German Science Foundation (DFG). This study is 
supported by the European Commission's FP7 Capacities Programme under the Grant 
Agreement number 312495.


\end{document}